\documentclass[useAMS,usenatbib]{mn2e}
\usepackage{txfonts}
\usepackage[T1]{fontenc}
\usepackage{ae,aecompl}

\usepackage{graphicx}
\usepackage{lscape}
\usepackage{amssymb}
\usepackage{rotating}
\usepackage{url}
\usepackage[british]{babel}

\newcommand{\mnras}{MNRAS}
\newcommand{\aap}{A\&A}
\newcommand{\apj}{ApJ}
\newcommand{\apjs}{ApJ}
\newcommand{\apjl}{ApJ}
\newcommand{\apss}{ASS}

\newcommand{\gppr}{\stackrel{>}{\scriptstyle \sim}}
\newcommand{\gappr}{\raisebox{-0.4ex}{$\gppr$}}
\newcommand{\lppr}{\stackrel{<}{\scriptstyle \sim}}
\newcommand{\lappr}{\raisebox{-0.4ex}{$\lppr$}}
\newcommand{\Mwd}{\mbox{$\mathrm{M_{WD}}$}}
\newcommand{\Msec}{\mbox{$\mathrm{M_{sec}}$}}
\newcommand{\Msun}{\mbox{$\mathrm{M_{\odot}}$}}
\newcommand{\Rsun}{\mbox{$\mathrm{R_{\odot}}$}}
\newcommand{\Porb}{\mbox{$\mathrm{P_{orb}}$}}
\newcommand{\Teff}{\mbox{$\mathrm{T_{eff}}$}}
\newcommand{\Ion}[2]{#1{\,\sc#2}}
\newcommand{\kms}{\mbox{$\mathrm{km\,s^{-1}}$}}

\title[Detached CVs are crossing the period gap]{Detached cataclysmic variables are crossing the orbital period gap}

\author[M. Zorotovic et al.]{
M. Zorotovic$^{1}$\thanks{E-mail: mzorotovic@dfa.uv.cl (MZ)},
M. R. Schreiber$^{1,2}$,
S. G. Parsons$^{1}$,
B. T. G\"ansicke$^{3}$,
\newauthor A. Hardy$^{1}$,
C. Agurto-Gangas$^{1,4}$, 
A. Nebot G{\'o}mez-Mor{\'a}n$^{5}$,
\newauthor A. Rebassa-Mansergas$^{6}$
and A. D. Schwope$^{7}$
\\
$^{1}$ Instituto de F\'isica y Astronom\'ia, Universidad de Valpara\'iso, Av. Gran Breta\~na 1111, Valpara\'iso, Chile\\
$^{2}$ Millennium Nucleus ``Protoplanetary Disks in ALMA Early Science'', Universidad de Valpara\'iso, Casilla 36-D, Santiago, Chile \\
$^{3}$ Department of Physics, University of Warwick, Coventry, CV4 7AL, UK\\
$^{4}$ Max-Planck-Institut fur extraterrestrische Physik, Giessenbachstrasse 1, 85748 Garching, Germany\\
$^{5}$ Observatoire Astronomique de Strasbourg, Universit\'e de Strasbourg, CNRS, UMR 7550, 11 rue de l'Universit\'e, F-67000 Strasbourg, France\\
$^{6}$ Departament de F\'isica, Universitat Polit\`ecnica de Catalunya, c/Esteve Terrades 5, 08860 Castelldefels, Spain\\
$^{7}$ Leibniz-Institut f\"ur Astrophysik Potsdam (AIP), An der Sternwarte 16, 14482 Potsdam, Germany
}

\date{Accepted 2016 January 27. Received 2016 January 26; in original form 2015 December 14}

\pubyear{2015}

\begin{document}
\label{firstpage}
\pagerange{\pageref{firstpage}--\pageref{lastpage}}
\maketitle

\begin{abstract}
A central hypothesis in the theory of cataclysmic variable (CV) evolution is the need to explain the observed lack of accreting systems 
in the $\simeq 2-3$\,h orbital period range, known as the \textit{period gap}. The standard model, disrupted magnetic braking (DMB), reproduces 
the gap by postulating that CVs transform into inconspicuous detached white dwarf (WD) plus main sequence (MS) systems, which no longer resemble CVs. 
However, observational evidence for this standard model is currently indirect and thus this scenario has attracted some criticism 
throughout the last decades. Here we perform a simple but exceptionally strong test of the existence of detached CVs (dCVs). If the theory is correct 
dCVs should produce a peak in the orbital period distribution of detached close binaries consisting of a WD and an M$4-$M$6$ secondary star.
We measured six new periods which brings the sample of such binaries with known periods below 10\,h to 52 systems. An increase of systems in 
the $\simeq 2-3$\,h orbital period range is observed. Comparing this result with binary population models we find that the observed 
peak can not be reproduced by PCEBs alone and that the existence of dCVs is needed to reproduce the observations. 
Also, the WD mass distribution in the gap shows evidence of two populations in this period range, i.e. PCEBs and more massive dCVs, which is 
not observed at longer periods. We therefore conclude that CVs are indeed crossing the gap as detached systems, which provides strong support for 
the DMB theory. 

\end{abstract}

% Select between one and six entries from the list of approved keywords.

\section{Introduction}

Cataclysmic variables (CVs) are close binaries in which a main-sequence (MS) donor transfers mass to a white dwarf (WD). The evolution of CVs is driven 
by angular momentum loss due to gravitational radiation (GR) and the much stronger magnetic braking (MB). The observed orbital period distribution of 
CVs has an apparent lack of systems in the $\simeq 2-3$\,h orbital period range, known as the \textit{period gap}. In order to explain this deficit, 
\citet{rappaportetal83-1} proposed a disrupted magnetic braking (DMB) scenario assuming that MB turns off when the donor star becomes fully convective 
at \Porb\,$\simeq 3$\,h. Systems above the gap are driven closer due to GR and efficient MB. Due to the strong mass transfer caused by MB the donors are 
driven out of thermal equilibrium. Once the donor star becomes fully convective, at the upper edge of the gap, MB stops or at least becomes inefficient. 
This causes a drop in the mass-transfer rate, which allows the donor star to relax to a radius which is smaller than its Roche lobe radius. The system detaches, 
mass transfer stops and it becomes a detached white dwarf plus main sequence (WD+MS) binary, evolving towards shorter periods only via GR. At \Porb\,$\simeq 2$\,h 
the Roche lobe has shrunk enough to restart mass transfer and the system appears again as a CV at the lower edge of the gap. 

The DMB scenario not only explains the period gap, but also agrees well with several other observed characteristics of CVs. The model adequately reproduces the 
higher accretion rates of systems above the period gap \citep{townsley+bildsten03-1,townsley+gaensicke09-1} and the larger radii of donor stars in CVs above 
the gap with respect to their MS radii \citep{kniggeetal11-1}. In addition, there is evidence for a discontinuity in the braking of single stars 
\citep{bouvier07-1,reiners+basri08-1} and/or a change in the field topology \citep{reiners+basri09-1,saundersetal09-1} around the fully convective boundary. 
Also in wide WD+MS binaries a significant increase in the activity fraction of M-dwarfs at the fully convective boundary has been observed 
\citep{rebassa-mansergasetal13-2}, which supports the idea that fully convective stars in wide binaries are not spun down as quickly as earlier M dwarfs. 
Finally, the prediction of the DMB scenario of a steep decrease of the number of post common envelope binaries (PCEBs) 
at the fully convective boundary \citep{politano+weiler06-1} is in agreement with the observations \citep{schreiberetal10-1}. 

However, these pieces of evidence supporting the standard theory of CV evolution based on the DMB scenario are rather indirect and the hypothesis that CVs are 
really crossing the gap as detached systems has been frequently challenged \citep[e.g.,][]{clemensetal98-1,andronovetal03-1,ivanova+taam03-1}. 
In addition, the standard scenario for CV evolution is facing several major problems, the most severe being that the predicted WD masses in CVs
are systematically smaller than the observed ones \citep{zorotovicetal11-1}. \citet{schreiberetal16-1} recently suggested a revision of the standard model
of CV evolution incorporating an empirical prescription for consequential angular momentum loss (CAML), i.e. angular momentum loss generated by mass transfer, 
and showed that the WD mass problem and several others can be solved if CAML is assumed to increase as a function of decreasing WD mass.  

A direct test of the main prediction of the standard scenario of CV evolution has been suggested by \citet{davisetal08-1}: 
if MB is disrupted at the upper boundary of the gap causing CVs to stop mass transfer, these systems should show up in population studies of detached WD+MS binaries. 
In particular, the deficit of CVs in the $\simeq 2-3$\,h orbital period range should imply an excess of short-period detached WD+MS binaries in the same period range. 
Observationally identifying this peak would provide clear evidence for the standard theory of CV evolution and may even allow to distinguish between the 
classical standard model and the revised version by \citet{schreiberetal16-1} as the latter predicts the CVs crossing the gap to contain more massive WDs. 

We here present the results of an observational search for close detached WD+MS binaries testing if the predicted peak at orbital periods of $\simeq 2-3$\,h exists. 
Comparing the observational results with those predicted by binary population models, we find that the existence of detached CVs (dCVs) is required to reproduce the
observations, and we therefore conclude that indeed CVs are crossing the gap as detached systems. In addition, as predicted by the revised model proposed by
\citet{schreiberetal16-1}, the observed WD mass distribution of detached systems in the $\simeq 2-3$\,h period range shows evidence for a combined
population of PCEBs and dCVs (more massive) in the period gap. 

\section{The spectral types of detached CVs}\label{sec:dCVs}

If the standard theory of CV evolution is correct and CVs are crossing the $\simeq 2-3$\,h gap as detached systems, a peak of systems should show up in the orbital 
period distribution of detached PCEBs with secondary stars of spectral types that are expected for dCVs. Detached CVs should have secondary stars with similar 
spectral types across the entire period gap, because MB is assumed to stop when the secondary star becomes fully convective and the mass remains constant while they are 
detached within the gap. Single M-dwarfs become fully convective at \Msec\,$\simeq 0.35$\,\Msun\, \citep{chabrier+baraffe97-1} which correspond to a spectral type of 
M$3-$M$4$ \citep{rebassa-mansergasetal07-1}. However, in most CVs above the period gap the spectral type of the secondary star is significantly later than the 
spectral type expected for a zero-age MS star with the same mass \citep[e.g.,][]{baraffe+kolb00-1}. 
The mass at which a mass-losing star becomes fully convective is smaller than for single stars or secondary stars in detached systems. Using observational constraints 
from a large sample of CVs, \citet{knigge06-1} find the fully convective boundary for CVs to be at \Msec\,$= 0.2\pm0.02$\,\Msun, which according to
\citet{rebassa-mansergasetal07-1} corresponds to a spectral type of $\sim$ M$6$. However, the exact mass at which a CV secondary star becomes fully convective will 
differ from system to system. For example, it is affected by the time the system spent as a CV before reaching the fully convective boundary. If it started mass 
transfer very close to the upper edge of the period gap, with the secondary being close to fully convective, the secondary star may be only slightly out of thermal 
equilibrium when becoming fully convective compared to a system with a longer mass transfer history. This implies that dCVs may cover a range of secondary masses 
with \Msec\,$\sim0.2-0.35$\,\Msun\, which roughly corresponds to spectral types later than $\sim$ M$4-$M$6$ according to \citet{rebassa-mansergasetal07-1}. 
This fits with the spectral type range for PCEBs that start mass transfer within the period gap if we use the mass spectral-type relation from 
\citet{rebassa-mansergasetal07-1} for detached systems. Therefore, we decided to search for the peak produced by dCVs in the orbital period distribution 
of a large and unbiased sample of close detached WD+MS binaries with secondaries of spectral type M$4-$M$6$ assuming an uncertainty of half a subclass.

However, we are aware of the fact that spectral types of dCVs as well as spectral-type mass and spectral-type radius relations are notoriously uncertain. 
Therefore we performed several tests moving the spectral-type range assumed for dCVs one class towards earlier/later spectral types and find that the conclusions 
of this paper remain identical.
 
\section{The observed sample}

\begin{table*}
\centering
\caption[]{Six SDSS WD+MS binaries with new orbital period measurements. Uncertainties in the periods are given in parentheses.}
\label{tab:newWDMS}
\begin{tabular}{llccccc}  
\hline
System        & \Porb & $\gamma$ & $K_2$  & \Teff & $\Mwd$ & Sp2 \\
              & [d]   & [km/s] & [km/s] & [K]   & [\Msun] &   \\
\hline
SDSSJ111459.93+092411.0 & 0.2102534(1)  & -8.2 $\pm$ 1.6 & 143.9 $\pm$ 2.1 & $10324\pm172$ & $0.610\pm0.115$ & M5      \\
SDSSJ113006.11-064715.9 & 0.3085042(7)  & 15.5 $\pm$ 3.6 & 120.1 $\pm$ 4.8 & $11139\pm192$ & $0.520\pm0.076$ & M5      \\
SDSSJ121928.05+161158.7 & 0.674080(1)   & -3.2 $\pm$ 0.9 & 153.8 $\pm$ 1.4 & $7123\pm103$  & $0.930\pm0.124$ & M6      \\
SDSSJ143017.22-024034.1 & 0.18140900(9) & -28.5 $\pm$ 1.7 & 167.4 $\pm$ 2.1 & $10802\pm436$ & $0.640\pm0.201$ & M5       \\
SDSSJ145238.12+204511.9 & 0.10621803(3) & -44.1 $\pm$ 1.1 & 356.5 $\pm$ 1.4 & $-$           & $\geq0.89$ & M4       \\
SDSSJ220848.32+003704.6 & 0.103351(9)   & 8.3 $\pm$ 0.7 & 228.4 $\pm$ 1.0 & $-$           & $\geq0.33$ & M5       \\
\hline
\end{tabular}
\end{table*}

In what follows we describe our observational sample. We also present the details of the observations, data reduction and period determination for six systems. 
We analysed the observed orbital period distribution and the possible biases that affect our sample.

\subsection{Systems from the SDSS PCEB survey}

Our observational sample is mostly based on the results of a large project we performed over the last decade. The Sloan Digital Sky Survey (SDSS) 
sample of spectroscopically identified WD+MS binary stars \citep{rebassa-mansergasetal12-2} contains 2\,316 systems up to data release 8. The majority ($\sim3/4$) 
of these systems are wide binaries that never underwent a CE event \citep{schreiberetal10-1,rebassa-mansergasetal11-1,nebot-gomez-moranetal11-1}. We carried out a 
radial velocity (RV) survey to identify the PCEBs within the SDSS sample \citep{schreiberetal10-1}, and measured their orbital periods to constrain theories of 
close binary evolution \citep{nebot-gomez-moranetal11-1}. The target selection during this large observational project was mostly determined by observing 
constraints and otherwise random, i.e. was mostly independent of the secondaries spectral type. Only in a few cases we targeted systems with a certain spectral type,
e.g. when we were trying to measure the increase of systems across the fully convective boundary we preferentially observed systems with M$2-$M$4$ secondary stars.

The close binaries discovered in the above described project containing M$4-$M$6$ secondary stars with orbital periods measured through RVs or from 
ellipsoidal/reflection effect (25) are complemented with 22 eclipsing systems identified by combining our spectroscopic WD+MS identification from SDSS with 
photometry from archival Catalina Sky Survey data \citep{drakeetal10-1, parsonsetal13-1, parsonsetal15-1}.  This way we established a sample of 47 PCEBs with an 
orbital period below 10\,h and companions with spectral types M$4-$M$6$. 

We also included in our sample 11 systems with earlier spectral types (M$2-$M$3$) and orbital periods below 10\,h, selected in the same way, as a control group. 
We do not expect to see any detached systems in the period gap for this control group, because PCEBs with companions in this spectral-type range should start 
mass transfer at longer periods. 

\subsection{VLT/FORS survey of dCVs}

\begin{figure*}
\includegraphics[width=\textwidth]{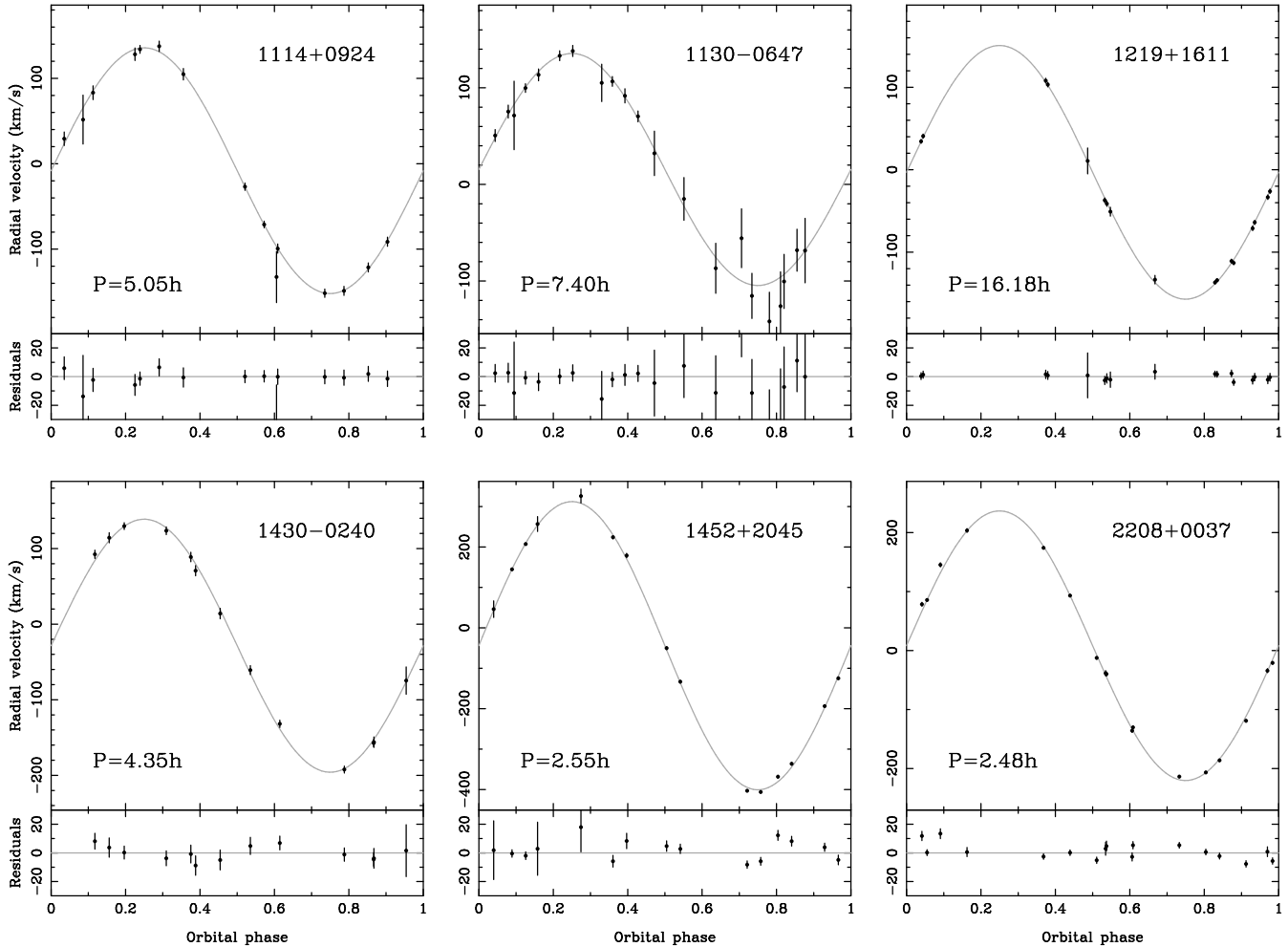}
\caption{Phase-folded RV curves for the six systems with periods presented in this paper. The grey line in each panel shows the sine fit to the data (see 
Table\,\ref{tab:newWDMS} for more details). The panel below each RV curve shows the residuals to this fit.}
\label{fig:rv}
\end{figure*}

To complement the sample that extracted from previous surveys, we carried out a dedicated search of close WD+MS systems with M$4-$M$6$ companions to search for 
dCVs crossing the gap. We measured 6 periods, five of them shorter than 10\,h. This brings our sample size to 52. In the following we describe the observations and data 
reduction.

We selected six systems from our catalogue of WD+MS binaries and observed them with the Very Large Telescope (VLT)
UT1 equipped with FORS2 \citep{appenzelleretal98-1} on the nights of 2014 May 16-18 
and 2015 July 2-4, in order to determine their orbital periods. We used the long slit mode with a 0.7" slit, 2x2 binning, the 1028z grism and the OG590 filter, 
resulting in a wavelength coverage of $7\,700-9\,500$\,{\AA} with a dispersion of 0.8\,\AA/pixel. The data were reduced using the standard ESO reduction pipeline. We also 
applied a telluric correction to the data using observations of the DQ WD GJ 440 taken at the start of each night. We measured the RV of the M dwarf in each 
spectrum by fitting the \Ion{Na}{i} absorption doublet at $\sim 8\,200$\,{\AA} with a combination of a straight line and two Gaussians of fixed separation, typically reaching a precision of 
5-10\,\kms\, in each individual spectrum. We then determined the orbital periods of the binaries by fitting a constant plus sine wave to the velocity measurements over 
a range of periods and computing the $\chi^2$ of the resulting fit. In Fig.~\ref{fig:rv} we show the phase-folded RV curves and corresponding fits for these systems. 
Table~\ref{tab:newWDMS} lists the results of these fits and the parameters of the systems, where the WD temperatures and masses are taken from 
\citet{rebassa-mansergasetal12-2} and 3D model corrections have been applied for systems with temperatures below $12\,000$\,K \citep{tremblayetal13-1}.
SDSS\,J1452$+$2045 and SDSS\,J2208$+$0037 show no absorption features from the WD in the SDSS spectra, meaning that the masses 
and temperatures can not be reliably determined. However the RV semi-amplitude can be used to determine a lower limit on the WD mass, which is provided in 
Table~\ref{tab:newWDMS}.

\subsection{Observed period distribution}

Our final sample contains 52 close WD+MS binaries with orbital periods \Porb\,$\leq 10$\,h and spectral types M$4-$M$6$. Their parameters as well as an explanation of how the close binary nature has been 
revealed and how the period has been measured are listed in Table~\ref{tab:all} in the Appendix.
The masses and temperatures of WDs cooler than $12\,000$\,K have been updated based on 3D model 
corrections \citep{tremblayetal13-1} except in some systems where the inclination is constrained by the eclipse which places a limit on the WD mass that is more accurate
than the mass estimated from the spectra. 

\begin{figure*}
\begin{center}
\includegraphics[angle=270,width=\columnwidth]{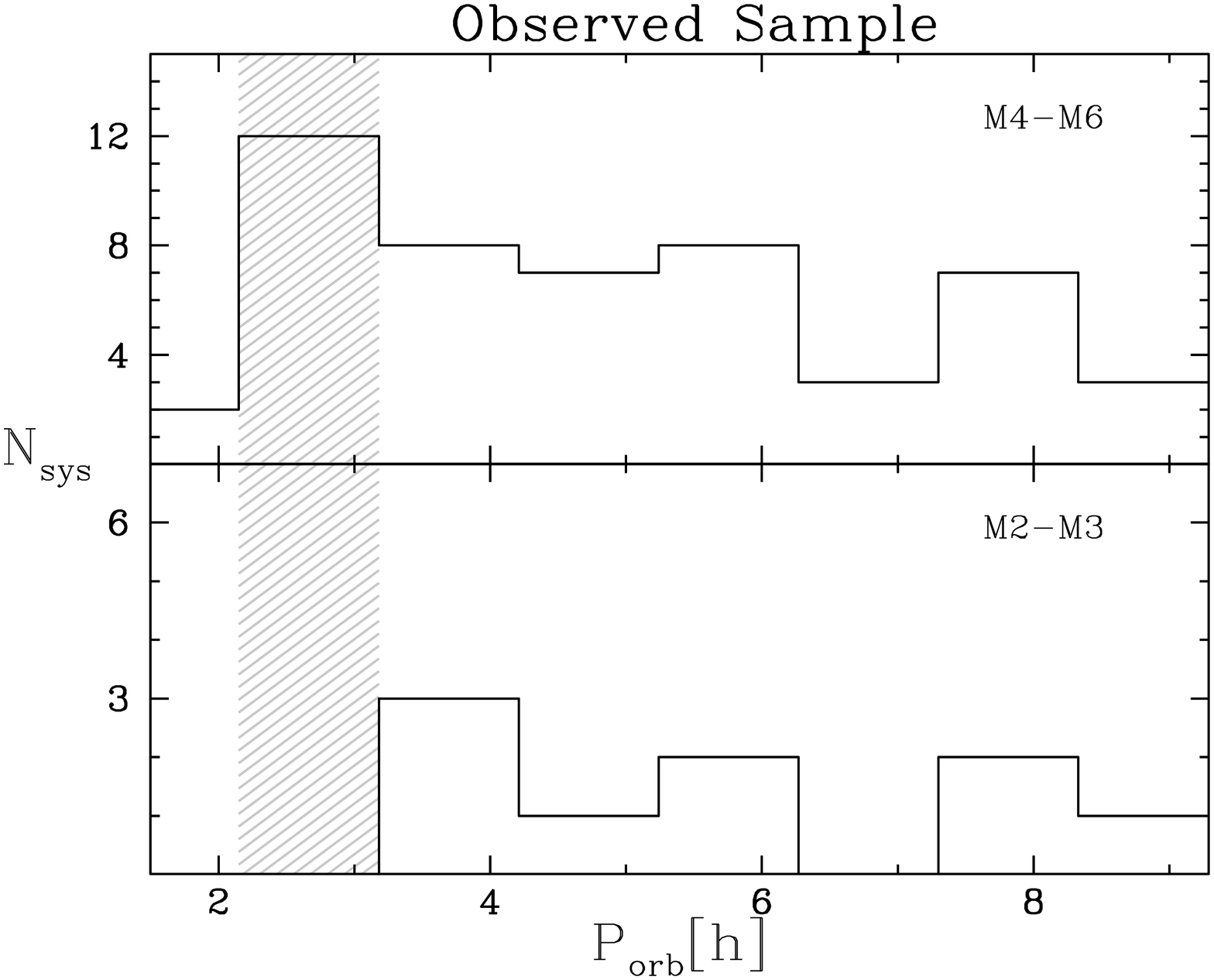}
\includegraphics[angle=270,width=\columnwidth]{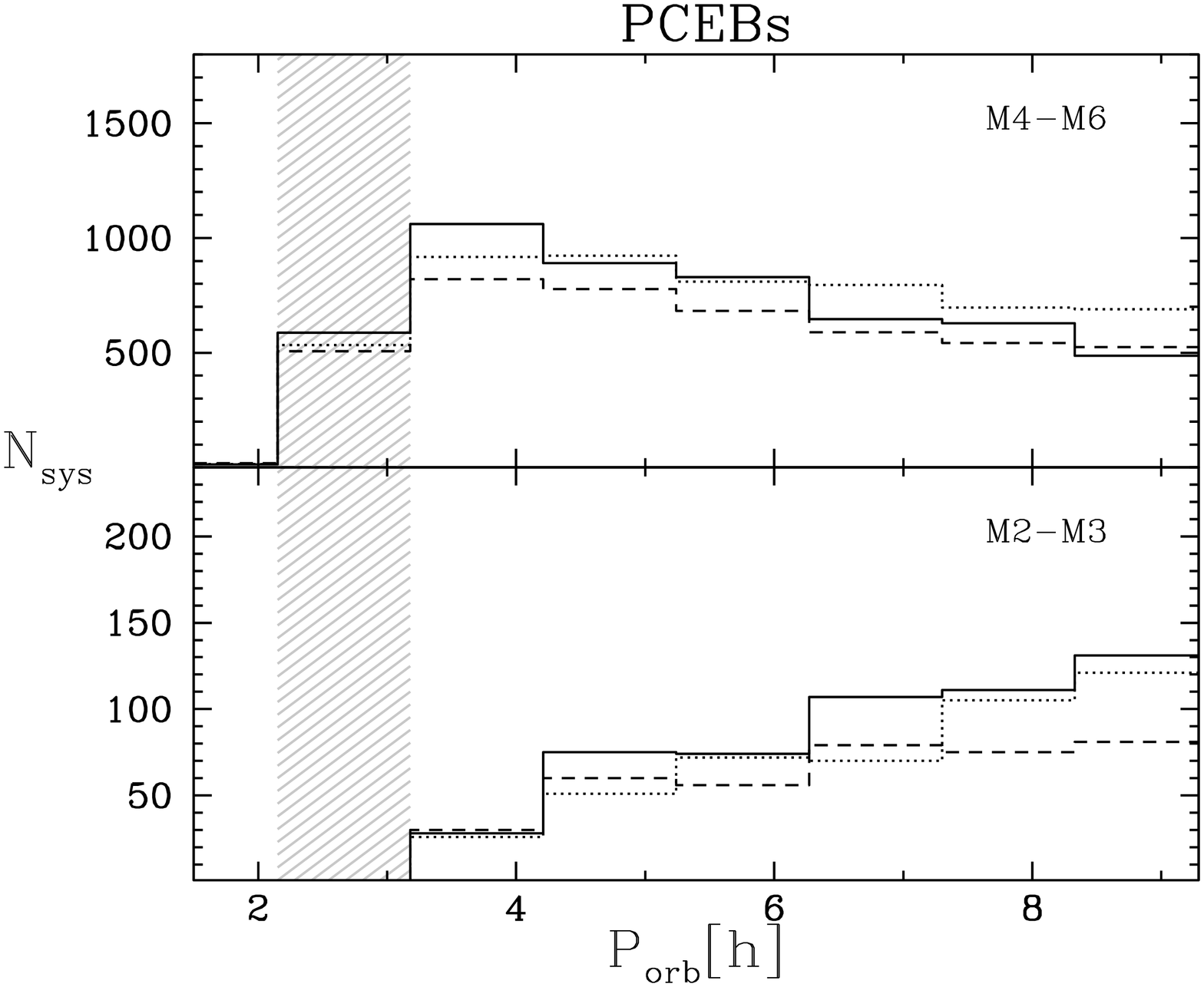}
\end{center}
\caption{\textit{Left:} Observed period distribution for detached WD+MS binaries. The upper and bottom panels show the distribution for different ranges of spectral types for 
the main-sequence companion. The corresponding ranges are labelled in the top right corner of each panel. The hatched area represents the location of the period gap 
($2.15-3.18$\,h, \citealt{knigge06-1}). \textit{Right:} Orbital period distribution for the simulated population of PCEBs with secondary stars in the 
corresponding spectral-type ranges. The solid line corresponds to the simulations with $\alpha_\mathrm{CE} = 0.25$, the dotted line to
$\alpha_\mathrm{CE} = 0.5$, and the dashed line to $\alpha_\mathrm{CE} = 1.0$.}
\label{fig:obs}
\end{figure*}

The left panel in Fig.~\ref{fig:obs} shows the observed orbital period distribution of close detached WD+MS binaries in our sample with secondary stars in the 
spectral-type range M$4-$M$6$ (top) and M$2-$M$3$ (bottom). The hatched area corresponds to the period gap according to \citet{knigge06-1}. The 
binning has been chosen to cover the whole gap in only one bin ($2.15-3.18$\,h). A peak can be observed at the position of the period gap for systems containing 
M$4-$M$6$ companions.  On the other hand, the period distribution of PCEBs with secondaries in the spectral-type range M$2-$M$3$ only contains systems with periods 
above the gap. This confirms that we have selected the correct spectral-type range to search for dCVs and that our results are not affected by the uncertainty of 
the spectral type of the secondary star, which is typically roughly half a subclass \citep{rebassa-mansergasetal07-1}. 

\subsection{Possible observational biases}

The close WD+MS binaries in our sample have been identified through RV variations or eclipses in their light curves. Both methods imply an observational bias towards 
short orbital periods that we have to consider before comparing the observed period distribution with the results of binary population models. 

Systems with shorter periods show larger RV variations and therefore their close nature is easier to determine. However, as shown in 
\citet[][ their Figure\,10]{nebot-gomez-moranetal11-1} the detection probability of close binarity only significantly decreases at periods longer than about one day. 
Therefore, for the majority of systems (35) in our sample which have been identified as close binaries through RV measurements, we can clearly exclude observational 
biases to affect our results. 

Eclipse light curves led to the discovery of the close binary nature in 17 systems in our sample. As shown by \citet{parsonsetal13-1}, the baseline and cadence of the 
archival Catalina data is typically good enough to detect eclipsing systems with orbital periods of about a day, so the detection probability again should not
affect our results. However, the detection probability is not the only possible bias towards shorter periods in the case of eclipses. The smaller the orbital period, 
the wider the range of inclinations that produce an eclipse. In other words, there is a larger fraction of eclipsing systems at shorter orbital periods. This has been shown in 
\citet[][ Figure\,4]{parsonsetal13-1}, where a comparison between the period distribution of all SDSS spectroscopically confirmed eclipsing PCEBs and all SDSS PCEBs 
from \citet{nebot-gomez-moranetal11-1} has been performed. To test whether the latter bias could affect our results, we investigated the fraction of eclipsing systems
in our observational sample and found that 40\% of the systems with M$4-$M$6$ companions are known to be eclipsers: 50\% of the systems in the period gap and 38\% above 
(see Table~\ref{tab:all} in the Appendix). This confirms that the fraction of eclipsing systems is larger in the bin with the shortest periods, although we can not exclude that 
this is caused by the low number of systems. Also, in some of these systems
their close nature was initially revealed from RV variations and their eclipsing nature was subsequently discovered. We found that the fraction of systems that were 
discovered to be close solely due to their eclipsing nature is similar in the gap and outside (33\% and 30\%, respectively). This means that the potential bias towards shorter
periods caused by close systems identified through eclipses is not important and can not be responsible for the peak observed at the position of the period
gap in the upper-left panel of Fig.~\ref{fig:obs}. 

While we can exclude that our sample is significantly biased with respect to the orbital period, the situation is different concerning the temperature of the WDs in 
our systems. If the WD is colder than $\sim 8\,000$\,K it becomes very difficult to measure its temperature from SDSS spectra, because no hydrogen absorption lines are
present, which leads to a clear bias against old systems. The two systems in our observed sample with WD temperatures significantly below $8\,000$\,K 
(SDSS\,J0138$-$0016 and SDSS\,J1210$+$3347) are eclipsing and the WD temperatures were determined from their colours. This bias against systems containing 
cold WDs has to be taken into account when comparing observed and simulated populations. 

Finally, given the importance of the WD mass for our understanding of CV evolution, we consider possible biases affecting this parameter. 
The RV method for identifying close binaries causes a bias towards systems with larger WD masses because, for a given secondary mass and orbital period, 
the velocity of the secondary increases with WD mass. This bias does not affect the relative distribution of WD masses as it is independent of the orbital period 
(i.e. each orbital period bin is equally biased). In the case of eclipsing systems, the identification probability is virtually independent of the WD mass.

\section{Binary population models}

The observed period distribution of close but detached WD+MS systems with secondary spectral types of M$4-$M$6$ shows a peak at the position of the orbital 
period gap. In order to evaluate if this peak provides evidence for CVs crossing the gap as detached system we performed Monte Carlo simulations of the population of 
WD+MS PCEBs and dCVs in the period gap. In what follows we describe the
details of our population models.  

\subsection{PCEBs}

An initial MS+MS binary population of $10^7$ systems was generated. We assumed the initial-mass function of \citet{kroupaetal93-1} in the range $0.8-9$\,\Msun\, for the 
distribution of primary masses plus a flat initial mass-ratio distribution \citep{sanaetal09-1} for secondary masses, with a lower limit of 
\Msec\,$= 0.05$\,\Msun. A flat distribution in $\log a$ ranging from $3$ to $10^4$\,\Rsun\, was used for the orbital separations \citep{popovaetal82-1, kouwenhovenetal09-1} 
and a constant star formation rate was assumed with an upper limit of $10$\,Gyrs.

As in \citet{schreiberetal16-1}, the systems were first evolved until the end of the CE phase using the binary-star evolution (BSE) code from \citet{hurleyetal02-1}. 
Three different values of the CE efficiency were considered: $\alpha_\mathrm{CE} = 0.25$, $\alpha_\mathrm{CE} = 0.5$, and $\alpha_\mathrm{CE} = 1.0$. The subsequent 
evolution of these zero-age PCEBs was performed with our own code. All zero-age PCEBs were evolved to their current periods assuming systemic angular momentum loss 
due to MB and GR (if \Msec\,$> 0.35$\,\Msun) or GR only (if \Msec\,$\leq 0.35$\,\Msun). The normalization factors for MB and GR are based on the observational constraints 
derived by \citet{kniggeetal11-1}. If a system filled its Roche lobe it was not considered as a PCEB any more. 

The spectral-type range of the MS star was converted into a mass range based on the relation presented in \citet{rebassa-mansergasetal07-1}. The range M$4-$M$6$ corresponds to masses for the 
companion in the range $0.17-0.35$\,\Msun, which is consistent with the mass
range used by \citet{davisetal08-1}. Furthermore, this corresponds to the mass range of secondary stars that will commence mass transfer within the gap. PCEBs in this mass 
range are assumed to evolve towards shorter periods due to GR only. The spectral-type range M$2-$M$3$ used for comparison corresponds to a mass range of 
$0.35-0.45$\,\Msun\, for the companion. These systems are brought into contact mainly due to MB and therefore evolve faster towards a second mass transfer phase. 
These systems will start the second mass transfer phase at periods above the gap and therefore we do not expect to see any such system within the gap.

\subsection{Cataclysmic Variables}

To estimate the impact of dCVs on the predicted population of close detached systems with secondary spectral types of M$4-$M$6$ we extended the binary population 
synthesis model described above by incorporating CV evolution following \citet{schreiberetal16-1}. Once the secondary star fills its Roche lobe it is inflated 
to a larger radius based on the Mass-Radius relation for CVs above the gap given by \citet{kniggeetal11-1}. We stop MB when the secondary star reaches 
$0.2$\,\Msun\, and the system becomes a dCV which evolves through the period gap only via GR. 

As shown in \citet{schreiberetal16-1}, the simulated population of CVs is strongly affected by the critical mass ratio that is assumed for having stable 
mass transfer. Apart from the intrinsic angular momentum loss due to GR and MB, consequential angular momentum loss (CAML), i.e. angular momentum loss due 
to mass transfer and mass loss during the nova eruptions, can play an important role. 
Two different models for CAML are simulated: the classical non-conservative
model for CAML (cCAML) where the change in angular momentum is given by
\begin{equation}
\frac{\mathrm{\dot{J}_{CAML}}}{\mathrm{J}}= \frac{\Msec^2}{\Mwd(\Mwd+\Msec)} \frac{\mathrm{\dot{M}_{sec}}}{\Msec}
\end{equation}
\citep[see e.g.][]{king+kolb95-1}, and an empirical CAML (eCAML) model given by
\begin{equation}
\frac{\mathrm{\dot{J}_{CAML}}}{\mathrm{J}}=\frac{0.35}{\Mwd} \frac{\mathrm{\dot{M}_{sec}}}{\Msec}
\end{equation}
\citep{schreiberetal16-1} that recently has shown to solve several problems between predictions and observations of CVs, 
especially the disagreement between observed and predicted WD masses. 
In the eCAML model we adjusted the normalization factors for MB and GR in order to obtain mass transfer rates in CVs that are consistent with the ones obtained with
the cCAML model. The factors we used are 0.43 for MB and 1.67 for GR (instead of 0.66 and 2.74, respectively, from \citealt{kniggeetal11-1}). This means that 
systems evolve slower towards shorter periods when there is no mass transfer. However, as the star formation rate is constant and we do not take into account old (cool) systems, these 
factors should not affect the orbital period distribution for the PCEB population. The spectral-type mass conversion was performed as in the case of PCEBs.

\section{Comparison with the observations}

To compare the simulated populations with the observations we excluded systems with old WDs that are too cool to be reliably detected 
through observations, because the observed sample is strongly biased against such systems. 
The detectability of a WD against a companion of the same spectral type depends mostly on the WD effective temperature and only little 
on its mass \citep{zorotovicetal11-1}. Therefore, applying a temperature limit of $8\,000$\,K to all our systems seems reasonable for comparing observed 
and simulated populations. We estimated the effective temperature of the WDs using the cooling tracks by 
\citet{althaus+benvenuto97-1}\footnote{\url{http://fcaglp.fcaglp.unlp.edu.ar/evolgroup/TRACKS/tracks_heliumcore_prev.html}} for helium-core WDs  
(if \Mwd\,$\lappr\,0.5$\,\Msun) and \citet{fontaineetal01-1}\footnote{\url{http://www.astro.umontreal.ca/~bergeron/CoolingModels}} for carbon/oxygen-core WDs 
(if \Mwd\,$\gappr\,0.5$\,\Msun). The temperatures of the WDs in PCEBs and dCVs were calculated in the same way. 
We note that the effective temperature of the WD in a dCV might be affected by compressional heating during the previous CV phase 
\citep{sion95-1,townsley+bildsten04-1,townsley+gaensicke09-1}. However, it is not clear how long it takes for the WD to cool down after mass transfer stops. 
If the time-scale is longer or comparable to the time a dCV spends in the gap,
i.e. if the effective temperature of a dCV is higher than the cooling temperature, 
the number of dCVs produced by the simulations can be slightly underestimated. 

We start our comparison by using PCEBs only to test if the peak observed at the position of the period gap for systems with M$4-$M$6$ companions can be 
reproduced without assuming dCVs crossing the gap. 

\begin{figure*}
\begin{center}
\includegraphics[angle=270,width=0.49\textwidth]{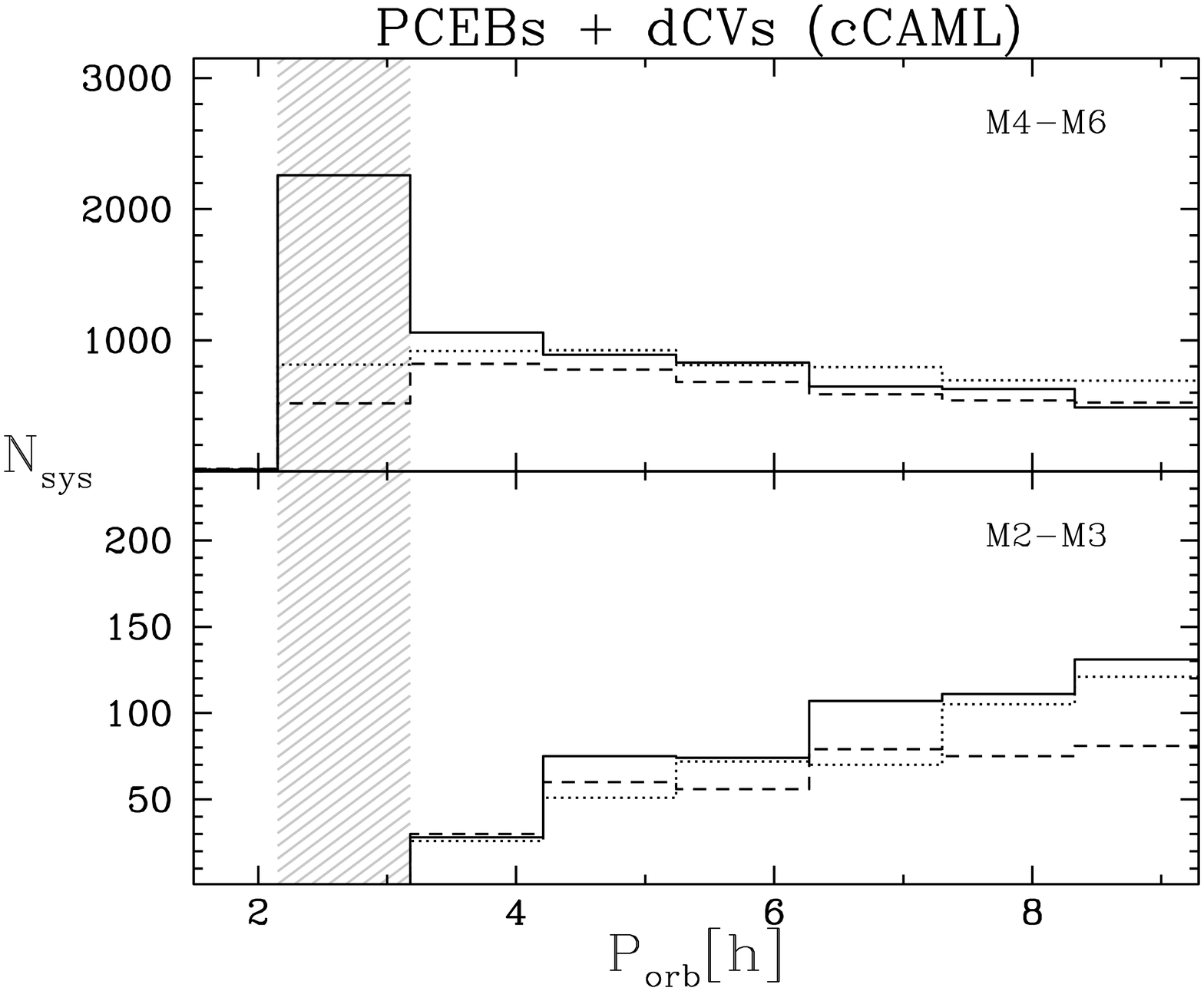}
\includegraphics[angle=270,width=0.49\textwidth]{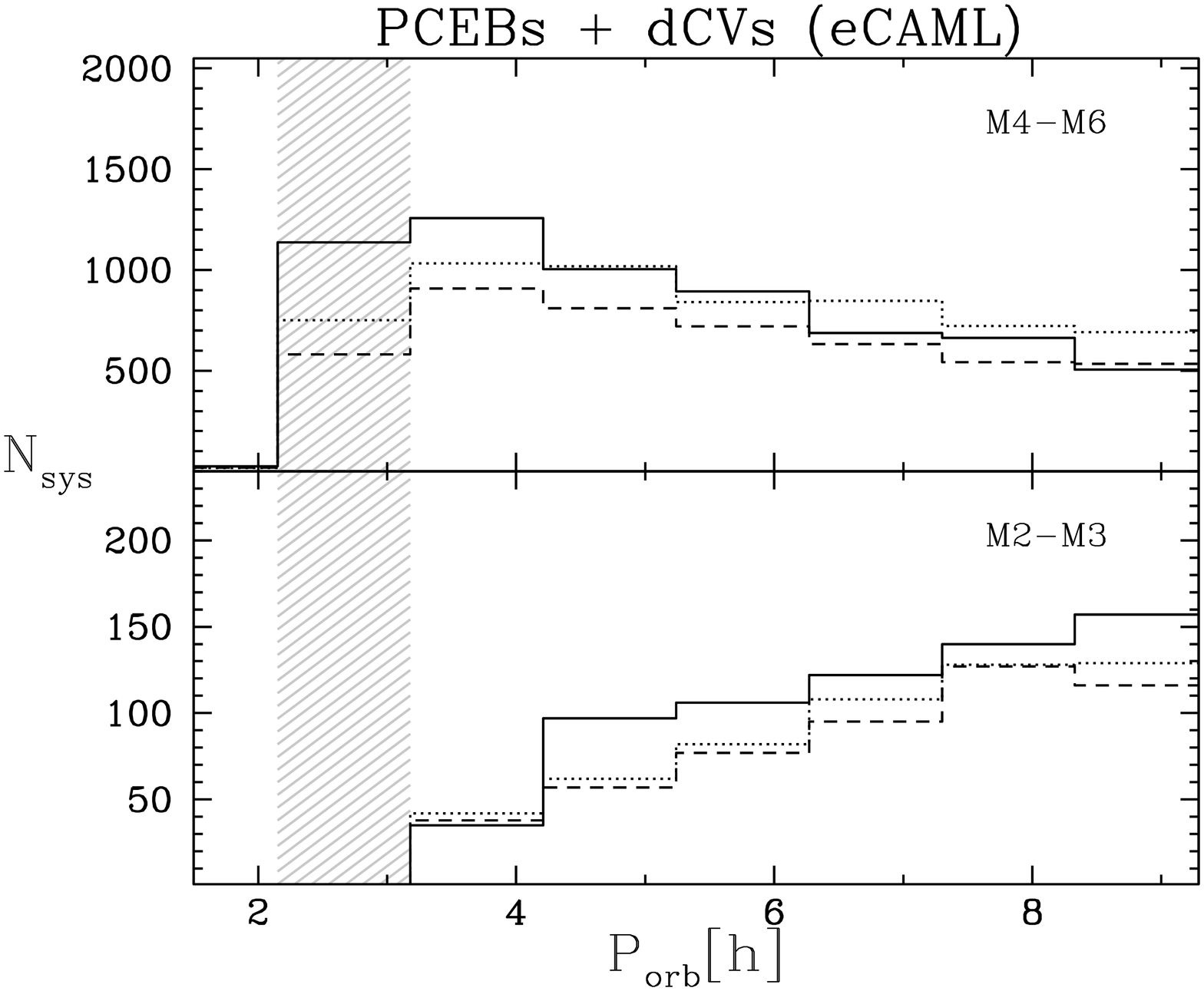}
\end{center}
\caption[]{Same as in Fig.~\ref{fig:obs} but including the population of dCVs in the period gap for our two models: cCAML (\textit{left}) and eCAML (\textit{right}). 
The different styles of lines correspond to different values for the CE efficiency: $\alpha_\mathrm{CE} = 0.25$ (solid), $\alpha_\mathrm{CE} = 0.5$ (dotted), and 
$\alpha_\mathrm{CE} = 1.0$ (dashed).}
\label{fig:GAPCVs}
\end{figure*}

\subsection{PCEBs }\label{sec:simP}

In the right panel of Fig.~\ref{fig:obs} we show the simulated orbital period distribution for PCEBs. The different lines 
correspond to different values of the CE efficiency. The upper panel, which contains systems with M$4-$M$6$ secondary stars, shows a trend to have more systems 
towards shorter periods with a drop in the period gap, independent of the CE efficiency parameter. In contrast, the bottom panel 
shows a decrease of systems with M$2-$M$3$ secondary stars towards shorter periods and none within the gap. 
This is expected because systems with earlier spectral types have more massive secondaries that fill their Roche lobes at longer periods.

While the observed and predicted distributions for systems with M$2-$M$3$ secondary stars agree quite well in not showing any system within the gap
(and keeping in mind that the observed sample is admittedly small), the observed peak at the period gap for systems 
with spectral types M$4-$M$6$ is in contrast to the drop expected from our simulations of the PCEB population if dCVs do not exist. 
In the next section we evaluate if this is better reproduced if we include dCVs.

\subsection{Including dCVs}\label{sec:simCV}

Figure~\ref{fig:GAPCVs} shows the simulated orbital period distribution for the combined population of PCEBs and dCVs in the same spectral-type ranges as in 
Fig.~\ref{fig:obs}. As expected, the distributions of the systems with M$2-$M$3$ secondaries (bottom panels) do not change because no detached systems 
with M$2-$M$3$ secondaries are produced by CV evolution. The difference between the distributions in the left and right panels in this range is purely 
due to the normalization factors for MB and GR assumed in each model. 

The predicted orbital period distributions for close detached systems with secondaries of spectral type M$4-$M$6$, however, are significantly affected. The number of 
systems in the orbital period range of the period gap is clearly increased. This effect is strongest for small values of $\alpha_\mathrm{CE}$ and stronger in the cCAML 
model than in the eCAML model. Comparing with the observed distribution (Fig.\,\ref{fig:obs}, left panel) it seems that especially models assuming small values for 
$\alpha_\mathrm{CE}$ provide a better agreement between theory and observations than is reachable with PCEBs alone.    

\begin{figure*}
\begin{center}
\includegraphics[width=0.32\textwidth]{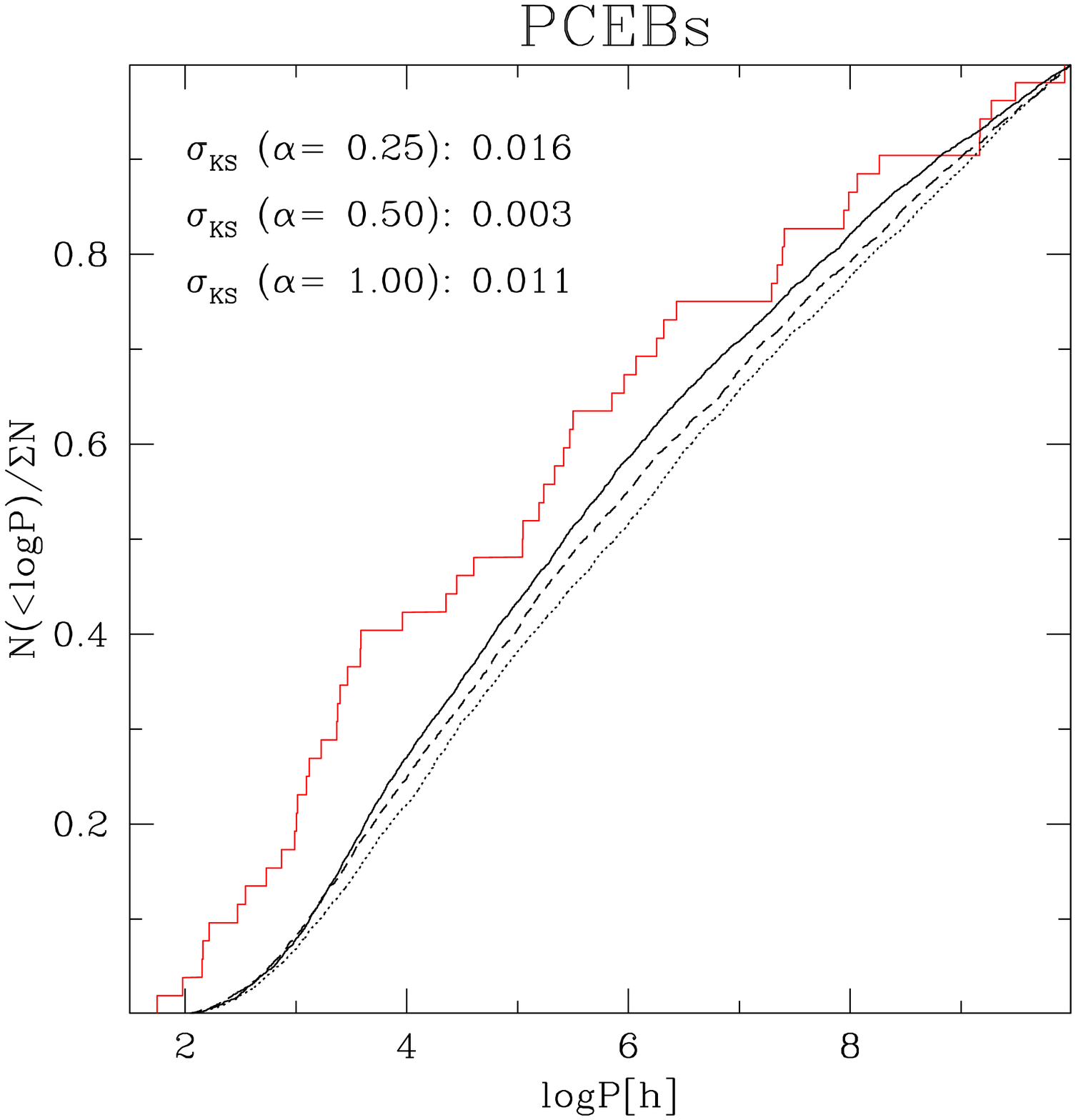}
\includegraphics[width=0.32\textwidth]{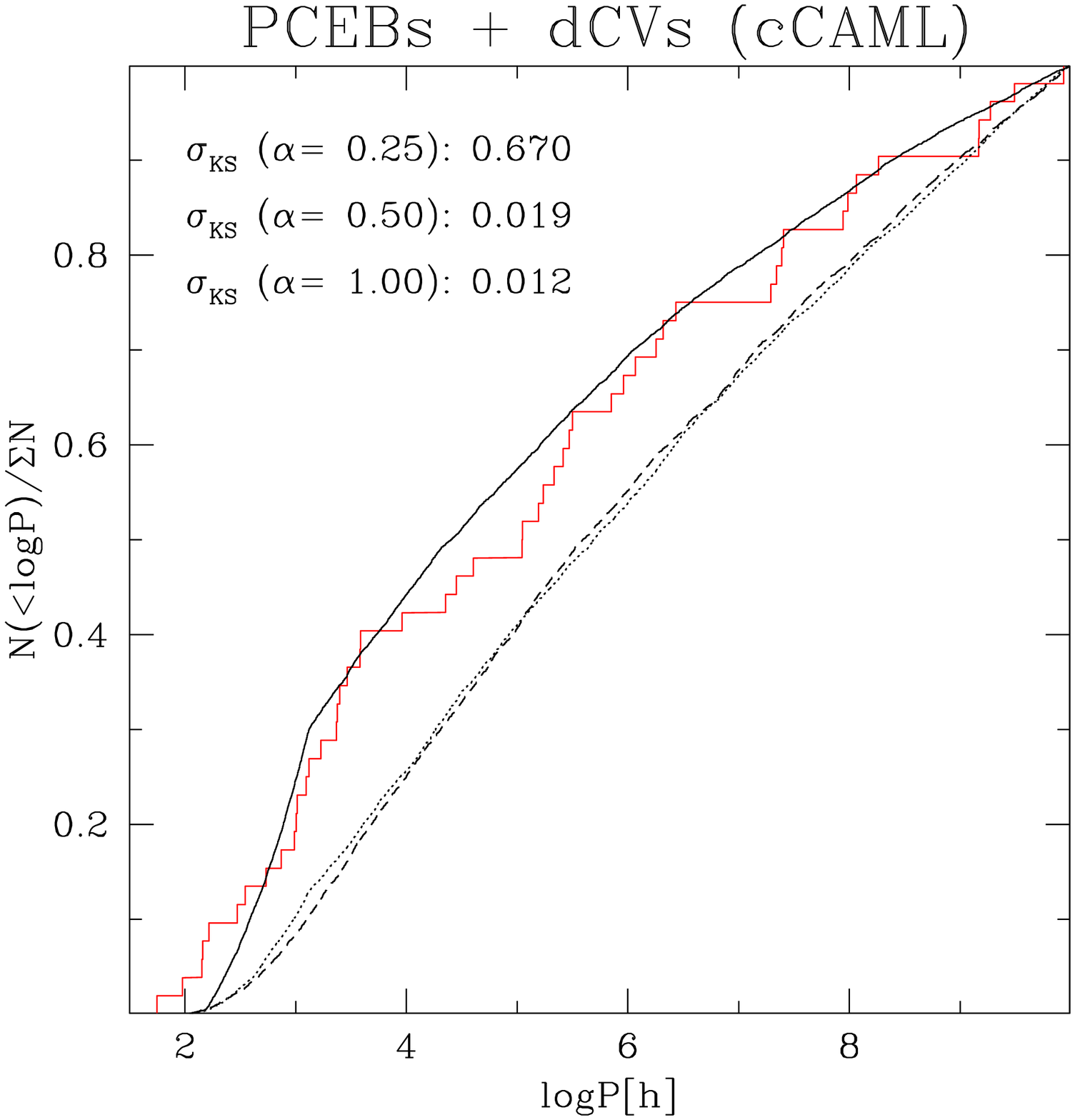}
\includegraphics[width=0.32\textwidth]{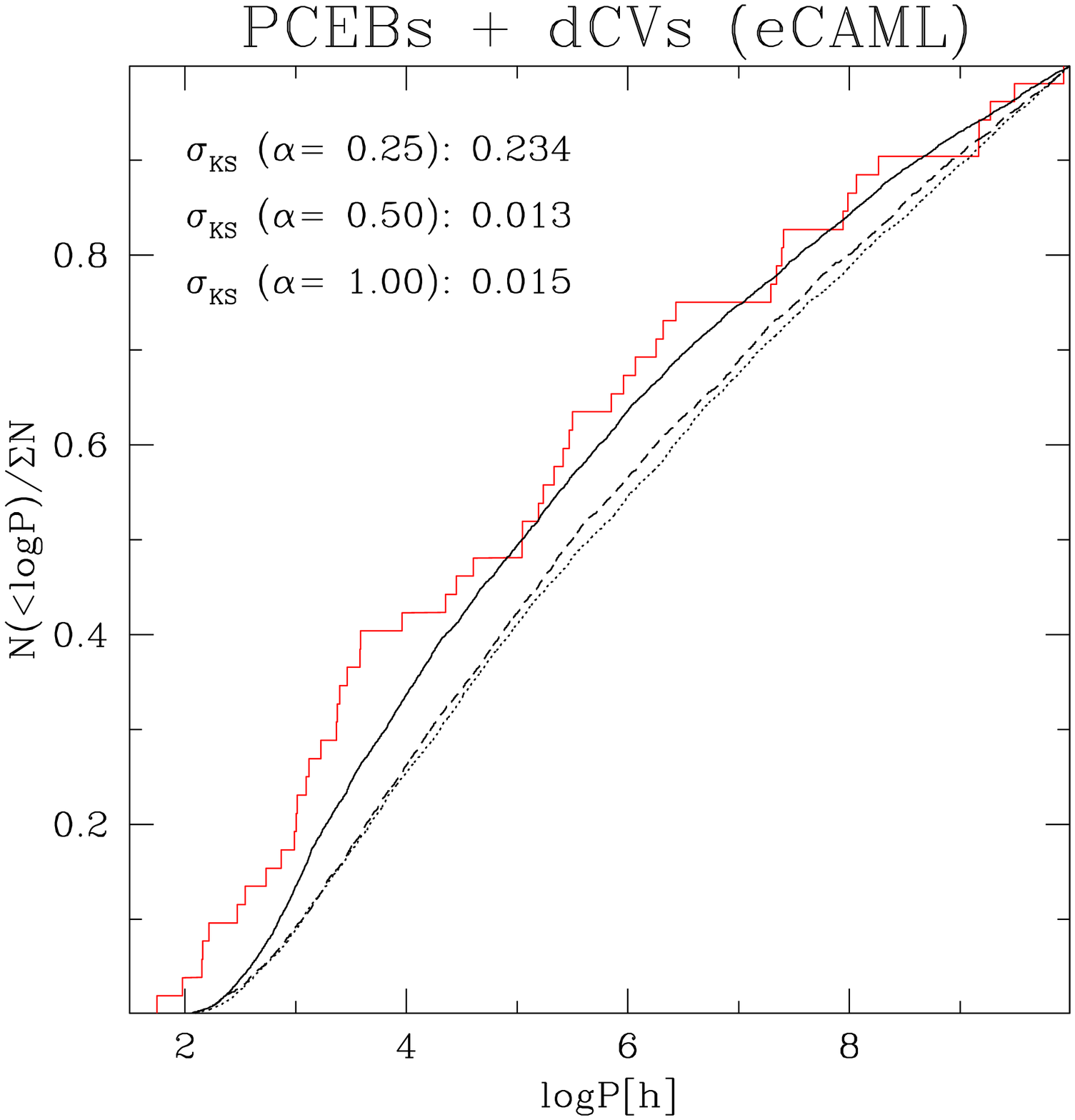}
\end{center}
\caption[]{Comparison between the observed (red) and simulated (black) cumulative distribution of orbital periods. Left panel: only PCEBs; middle panel: PCEBs + dCVs 
from the cCAML model; right panel: PCEBs + dCVs from the eCAML model. The different styles of lines in each panel correspond to different CE efficiencies: 
$\alpha_\mathrm{CE} = 0.25$ (solid line), $\alpha_\mathrm{CE} = 0.5$ (dotted line) and $\alpha_\mathrm{CE} = 1.0$ (dashed line). The values in the upper left corner 
of each panel are the KS probabilities.}
\label{fig:KS}
\end{figure*}

However, given the still relatively low number of systems in our sample, we need to carefully investigate whether this apparent improvement provides statistically robust 
evidence for the existence of dCVs crossing the gap. To that end, we performed a Kolmogorov-Smirnov (KS) test between the observed and simulated period distributions. 
Figure~\ref{fig:KS} shows the cumulative distributions of orbital periods for our simulations (black) and the observed systems (red) with MS companions in the spectral-type 
range M$4-$M$6$. The left panel compares the observational sample with our PCEB simulations (without dCVs), middle shows the comparison with PCEBs + dCVs from the 
cCAML model, and the right-hand panel compares observations with PCEBs + dCVs from the eCAML model. Different styles of lines correspond to the three different CE 
efficiencies used in the simulations. The KS probabilities are also listed in the upper left corner of each panel. According to the KS test, the cumulative distribution 
of observed systems and PCEBs is different with a confident level of at least $98.4\%$ depending on the value of the CE efficiency that is assumed. The corresponding KS 
probability is less than 0.02 and we therefore conclude that the two samples are different. In the case of PCEBs + dCVs, both models 
show larger KS probabilities when a small CE efficiency is assumed ($\alpha_\mathrm{CE} = 0.25$). For the cCAML model the KS probability is 0.670 while for the eCAML it 
is 0.234. Based on these values, we can not exclude any of the two models. However, the probabilities drop dramatically if we use larger values for the CE efficiency and 
all the models with $\alpha_\mathrm{CE} = 0.5$ or $\alpha_\mathrm{CE} = 1.0$ can be rejected with a confidence level larger than $98\%$. This is consistent with recent
studies that show that low values of $\alpha_\mathrm{CE}$ seem to work best for PCEBs with M-dwarf secondaries 
\citep[e.g.,][]{zorotovicetal10-1,toonen+nelemans13-1,camachoetal2014-1}. 
We also performed the KS tests for the simulated and observed systems with secondary stars in the spectral-type range M$2-$M$3$. Comparing with the predicted 
PCEB sample the KS probabilities are larger than 0.1 for both CAML models and for the three values of the CE efficiency assumed in our simulations. This means that
there are no statistically significant differences in the two distributions, i.e. observations and predictions agree.

\section{Disentangling detached CVs and PCEBs}

In the previous section we showed that PCEBs alone can not be responsible for the observed peak at the position of the period gap in the observed period distribution of 
detached close WD+MS systems with secondaries of spectral type M$4-$M$6$. 
If there were only PCEBs, one should expect to see a drop in the number of systems in this bin, because PCEBs with secondary stars in this 
spectral type range fill their Roche lobes within the gap. The inclusion of dCVs in the gap can reproduce the observed peak, and the size of the expected peak depends 
strongly on the model that we assume for CAML and on the efficiency for CE ejection. Only with a small value for $\alpha_\mathrm{CE}$ does the KS test show no significant 
differences between the observed and simulated distributions with dCVs, in agreement with \citet{zorotovicetal10-1}. Based on the current data, we can not decide which 
of the two models for CAML we tested should be preferred as both models produce reasonable agreement with the observations. However, our results clearly show that CVs 
are crossing the gap as detached systems, which provides further evidence for the DMB model. The question that immediately arises from this result is: 
is there a way of observationally distinguishing a normal PCEB from a dCV in the period gap? While the secondary stars of dCVs should be indistinguishable from those of PCEBs with 
M$4-$M$6$ secondaries, the WD parameter distributions may provide new insights. 

\subsection{WD masses}

\begin{figure*}
\begin{center}
\includegraphics[angle=270,width=0.7\textwidth]{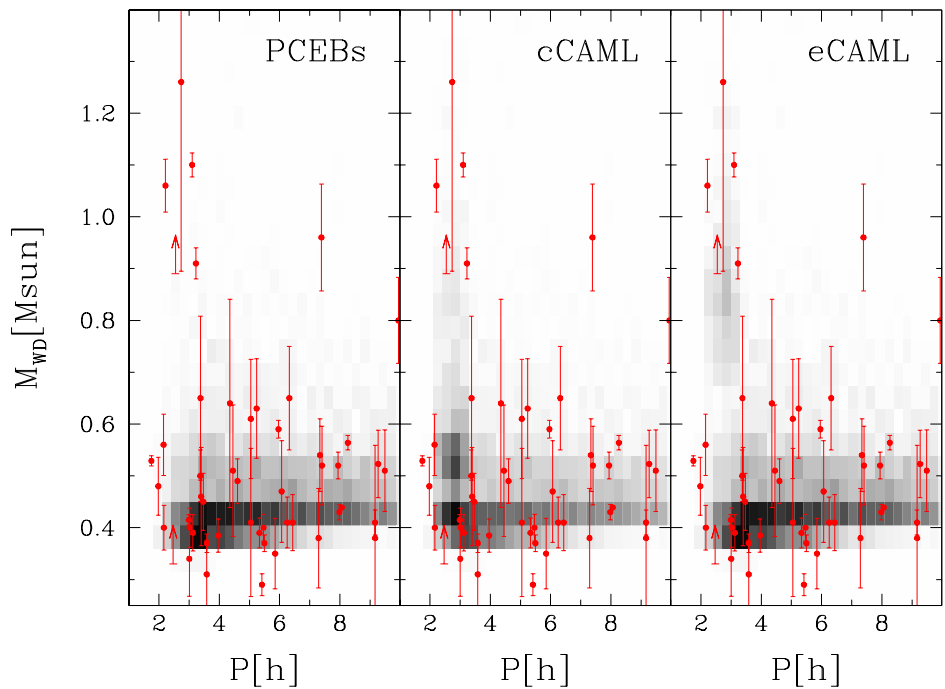}
\end{center}
\caption[]{Relation between WD mass and orbital period. The intensity of the grey scale represents the density of simulated objects in each bin, on a linear scale, and 
normalized to one for the bin that contains most systems. The red dots are the observed systems with available WD masses. The two vertical arrows correspond to 
SDSS\,J1452$+$2045 and SDSS\,J2208$+$0037 for which a lower limit for the WD mass has been calculated based on the RV semi-amplitude.}
\label{fig:Mwd}
\end{figure*}

As shown in \citet{zorotovicetal11-1}, the WD mass distribution of CVs and PCEBs is very different. WDs in CVs are, on average, more massive and there is a lack of 
low-mass WDs (helium-core WDs). If some of the systems within the period gap are in fact dCVs, one should expect to have a larger average WD mass in this period range than 
at longer periods.

In Figure~\ref{fig:Mwd} we show the distribution of WD masses and orbital periods for the observed systems with measured WD mass (red dots) and for our simulations 
assuming $\alpha_\mathrm{CE} = 0.25$ (grey scale density). In the simulation that only includes PCEBs (left panel) the systems are concentrated at low-mass 
WDs in all period ranges, exhibiting a single peak (at $\sim 0.40-0.45$\,\Msun) and a continuous decrement of systems towards more massive WDs. On the other hand, in our two simulations with dCVs 
a second population is clearly visible in the $\simeq 2-3$\,h orbital period range. In the case of the cCAML model (middle panel), a second peak is evident at 
$\sim 0.5-0.6$\,\Msun\, and systems with more massive WDs become more frequent at these periods. In the eCAML model the second peak is not as pronounced as in the cCAML 
model, but occurs at higher masses ($\sim 0.8-0.9$\,\Msun). This is because the eCAML model is more restrictive than the cCAML model with respect to the stability limits 
for mass transfer and therefore produces less CVs (and subsequently less dCVs) but with higher WD masses, which is more consistent with the observed WD mass distribution 
of CVs.

From the observational sample we see that the population of systems with massive WDs ($>0.8$\,\Msun) is concentrated at the location of the period gap. 
We obtain an average WD mass of $0.66\pm0.12$\,\Msun\, in the gap and $0.50\pm0.02$\,\Msun\, outside the gap, with standard deviations of $0.35$\,\Msun\, and $0.15$\,\Msun\, respectively, 
for the systems with M$4-$M$6$ companions. This is consistent with having some dCVs with massive WDs in the period gap and seems to provide further support for the eCAML model. 
However, the tendency of having high-mass WDs in the gap needs to be interpreted with caution because of two reasons: first, our sample is too small to provide a 
statistically significant result. Second, it has been previously found that some of the masses derived from spectra may overestimate significantly the true value,
especially if the spectrum is dominated by the MS star component \citep{parsonsetal13-1}. 
This is almost certainly the case for SDSS\,J0052$-$0053, the system in the gap with the most massive WD ($1.26$\,\Msun). However, 
for two other gap systems with $\Mwd>1$\,\Msun\, the WD is clearly visible in their SDSS spectra, meaning that these systems quite likely contain massive 
WDs. One of these systems, SDSS\,J1013$+$2724, is in fact an eclipsing system and \citet{parsonsetal15-1} noted that the sharp ingress and egress eclipse features are in 
agreement with a small (hence massive) WD. The large RV semi-amplitude of SDSS\,J1452$+$2045 (one of the new systems presented in this paper) also places a lower limit 
on the mass of this DC WD of $0.89$\,\Msun. In summary, there seems to be an excess of systems containing fairly massive WDs with periods in the gap and these may well be 
dCVs crossing the gap.

\subsection{WD effective temperatures}

A second possibility to distinguish dCVs and PCEBs might be the effective temperature of the WD. In comparison with CVs above the period gap, dCVs should be cooler because 
CVs suffer from compressional heating of the outer layers \citep{sion95-1}. 
On the other hand, dCVs should be hotter than PCEBs in the same orbital period range, because the initial mass of the secondary star must have been higher than the limit for 
non-fully convective secondaries (\Msec\,$\gappr\,0.35$\,\Msun) in order to start mass transfer above the gap. This means that angular momentum loss after the CE phase was 
mainly driven by MB. PCEBs in the orbital period range of the period gap, however, need to have less massive secondaries in order to still be detached systems in this period 
range ($\simeq 2-3$\,h). This means that after the CE phase they become closer only due to GR and therefore they evolve slower towards shorter periods. This effect, however, 
might be compensated by the fact that systems with more massive companions tend to emerge from the CE phase at slightly longer periods 
\citep[e.g.,][]{zorotovicetal11-2,zorotovicetal14-1}. Which of the two effects is stronger is uncertain because it depends on, e.g., the initial orbital period, the star 
formation rate, the CE efficiency or the strength of MB and GR.

\begin{figure*}
\begin{center}
\includegraphics[angle=270,width=0.7\textwidth]{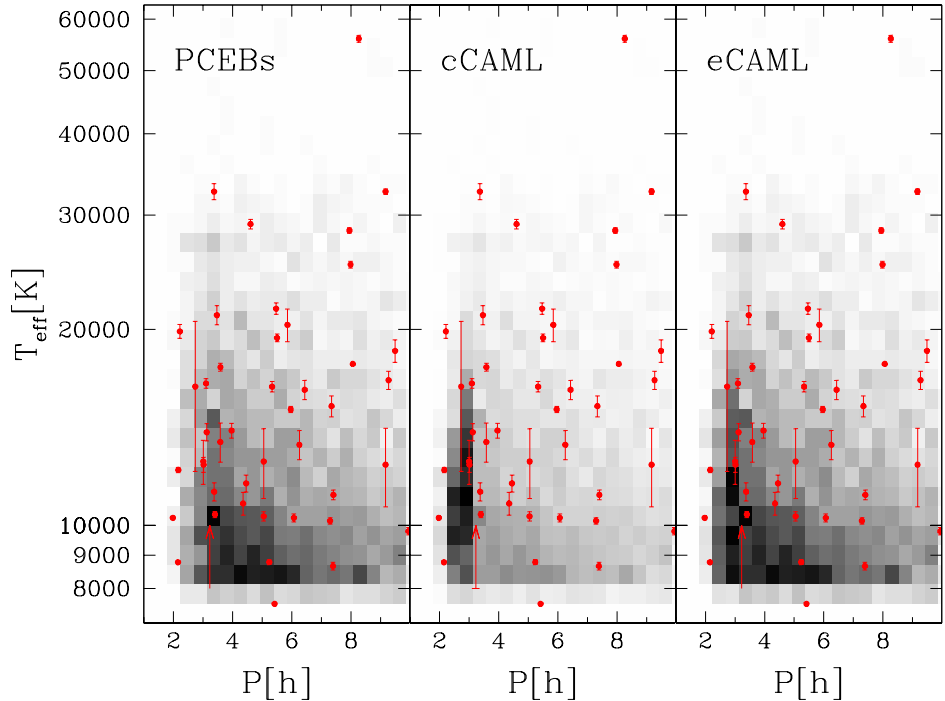}
\end{center}
\caption[]{Relation between the WD effective temperature and orbital period. The intensity of the grey scale for the simulations means the same as in Fig.\,\ref{fig:Mwd}. 
The red dots are the observed systems with available WD temperatures.}
\label{fig:teff}
\end{figure*}

Figure~\ref{fig:teff} shows the relation between WD effective temperature and orbital period for our simulations with $\alpha_\mathrm{CE} = 0.25$ (grey scale density plot) 
and for the observed systems with available WD temperatures (red dots). The two observed systems with the lowest temperatures are not represented in this figure because 
our simulations exclude cold ($<8\,000$\,K) WDs. The average temperature seems to increase towards shorter periods in all our models and there is no distinctive behaviour at periods 
corresponding to the period gap. The simulation that only includes PCEBs (left panel) looks almost identical to the one with dCVs from the eCAML model (right panel) while 
the cCAML model predicts a small increase of the number of hotter WDs in the orbital period range of the gap. This is because this model produces the largest fraction of 
dCVs compared to PCEBs at these periods. However, in general the predicted distributions of WD temperatures are not significantly different and, in agreement with 
this, the observed WD temperatures do not show any significant tendency
either. The observed WD effective temperature average is $\sim 13\,000\pm1\,300$\,K in the gap and $\sim 17\,000\pm 1\,600$\,K, outside the gap, with standard 
deviations of $\sim 4\,000$\,K and $\sim 9\,500$\,K respectively. The dispersion in both, observations and simulations, is substantial. We therefore conclude that the WD temperature is 
not a suitable parameter to identify dCVs within the period gap. 

\section{Conclusion}

We have measured six new periods of close detached WD+MS binaries with secondary stars in the spectral type range M$4-$M$6$, which should correspond to the spectral type range 
of secondary stars in detached CVs crossing the orbital period gap. These new measurements bring the sample of such binaries with measured orbital 
period to 52 systems. A clear peak in the orbital period distribution can be observed at the position of the orbital period gap, in agreement with the predictions from the 
disrupted magnetic braking model. Comparing the observed period distribution with the results of binary population models we find that this peak can not be explained without assuming 
that CVs are crossing the gap as detached systems. Therefore we conclude that indeed CVs become detached binaries at the upper edge of the period gap, which supports the idea 
that magnetic braking becomes inefficient at the fully convective boundary.

We also see clear signs of a different WD mass distribution in the gap and at longer periods. The WD mass distribution of systems within the gap shows a second peak at larger 
masses which is consistent with having two populations in this period range, i.e. normal PCEBs and the more massive detached CVs crossing the gap, in agreement with the model recently 
proposed by \citet{schreiberetal16-1}.  

\section*{Acknowledgements}
We thank Fondecyt for their support under the grants 3130559 (MZ), 1141269 (MRS), and 3140585 (SGP). 
The research leading to these results has received funding from the European Research Council under the European Union's Seventh Framework Programme (FP/2007-2013) / ERC Grant Agreement n. 320964 (WDTracer).
This research was also partially funded by MINECO grant AYA2014-59084-P and by the AGAUR (ARM).
The results presented in this paper are based on observations collected at the European Southern Observatory under programme IDs 093.D-0441 and 095.D-0739.

%%%%%%%%%%%%%%%%% APPENDICES %%%%%%%%%%%%%%%%%%%%%

\appendix

\section{Observational Sample}

\begin{table*}
\centering
\caption[]{Sample of detached WD+MS systems with M$4-$M$6$ companions used in this work. Systems in the period gap are highlighted in boldface.
The sixth column details how the close binary nature and period was determined with the following meaning: 
RV - found via RV variations, period measured from velocities;
RV$>$ELL - found via RV variations, period measured from ellipsoidal/reflection;
RV$>$ECL - found via RV variations, period measured from eclipses;
ECL - found via eclipses, period measured from eclipses.
\textit{References}.
~(1) \citet{rebassa-mansergasetal08-1}, ~(2) \citet{rebassa-mansergasetal12-2}
~(3) \citet{pyrzasetal09-1}, ~(4) \citet{parsonsetal12-2},
~(5) \citet{nebot-gomez-moranetal11-1}, ~(6) \citet{parsonsetal13-2}, ~(7) \citet{drakeetal10-1},
~(8) \citet{parsonsetal13-1}, ~(9) \citet{parsonsetal15-1}, ~(10) \citet{schreiberetal08-1},
~(11) This paper, ~(12) \citet{pyrzasetal12-1}, ~(13) \citet{nebot-gomez-moranetal09-1},
~(14) \citet{parsonsetal12-1}, ~(15) \citet{greenetal78-1}.\\
}
\label{tab:all} 
\begin{tabular}{lccccll}  
\hline
System  		&  \Porb   & Sp2 & \Teff & $\Mwd $ & Method & References\\
     			&  [h]     &     & [K]   & [\Msun] &        &  \\
\hline
\textbf{SDSSJ005245.11$-$005337.2} & \textbf{2.735(2)} & \textbf{4.0} & \textbf{16\,340$\pm$4\,240} & \textbf{1.260$\pm$0.365} & \textbf{RV}     & \textbf{1,2}\\
SDSSJ011009.09$+$132616.1 & 7.984495(3) & 4.0 & 25\,167$\pm$296 & 0.430$\pm$0.015 & RV$>$ECL & 3,2\\
SDSSJ013851.54$-$001621.6 & 1.7463576(5) & 5.0 & $3\,570_{-80}^{+110}$  & 0.529$\pm$0.010 & ECL    & 4\\
\textbf{SDSSJ015225.38$-$005808.5} & \textbf{2.15195(1)} & \textbf{6.0} & \textbf{8\,773$\pm$25}  & \textbf{0.560$\pm$0.059} & \textbf{RV}     & \textbf{5}\\
SDSSJ030308.36$+$005443.7 & 3.226505(1) & 4.5 & $<$8\,000    & 0.910$\pm$0.030 & RV$>$ECL & 3,6\\
SDSSJ031404.98$-$011136.6 & 6.32(2) & 4.0 & --    & 0.650$\pm$0.100 & RV     & 1\\
SDSSJ032038.72$-$063822.9 & 3.375(2) & 5.0 & 11\,264$\pm$361 & 0.650$\pm$0.158 & RV     & 5\\
SDSSJ083354.84$+$070240.1 & 7.34(2) & 4.0 & 15\,246$\pm$560 & 0.540$\pm$0.070 & RV     & 5\\
\textbf{SDSSJ083845.86$+$191416.5} & \textbf{3.122694(9)} & \textbf{5.0} & \textbf{13\,904$\pm$424} & \textbf{0.390$\pm$0.035} & \textbf{ECL}    & \textbf{7,2}\\
SDSSJ090812.04$+$060421.2 & 3.58652(6) & 4.0 & 17\,505$\pm$242 & 0.370$\pm$0.018 & ECL    & 7,2\\
SDSSJ093947.95$+$325807.3 & 7.943750(5) & 4.0 & 28\,398$\pm$278 & 0.520$\pm$0.026 & ECL    & 7,2\\
SDSSJ094634.49$+$203003.4 & 6.068669268(1) & 5.0 & 10\,268$\pm$141 & 0.470$\pm$0.098 & ECL    & 8,2\\
SDSSJ094913.37$+$032254.5 & 9.49(2) & 4.0 & 18\,542$\pm$737 & 0.510$\pm$0.079 & RV     & 5\\
\textbf{SDSSJ101356.32$+$272410.6} & \textbf{3.0969691(1)} & \textbf{4.0} & \textbf{16\,526$\pm$277} & \textbf{1.100$\pm$0.023} & \textbf{ECL}    & \textbf{9,2}\\
SDSSJ102102.25$+$174439.9 & 3.368617752(2) & 4.0 & 32\,595$\pm$928 & 0.500$\pm$0.050 & ECL    & 8,2\\
SDSSJ104738.24$+$052320.3 &9.17(2) & 5.0 & 12\,392$\pm$1715 & 0.380$\pm$0.179 & RV     & 10,2\\
\textbf{SDSSJ105756.93$+$130703.5} & \textbf{3.00389076(1)} & \textbf{5.0} & \textbf{12\,536$\pm$978} & \textbf{0.340$\pm$0.072} & \textbf{ECL}    & \textbf{8,2}\\
SDSSJ111459.93$+$092411.0 & 5.0460816(2) & 5.0 & 10\,324$\pm$172 & 0.610$\pm$0.115 & RV     & 11,2\\
SDSSJ113006.11$-$064715.9 & 7.40410(2) & 5.0 & 11\,139$\pm$192 & 0.520$\pm$0.076 & RV     & 11,2\\
SDSSJ114312.57$+$000926.5 & 9.27(3) & 4.0 & 16\,719$\pm$534 & 0.523$\pm$0.065 & RV     & 5\\
SDSSJ115156.94$-$000725.4 & 3.399(3) & 6.0 & 10\,395$\pm$114 & 0.460$\pm$0.095 & RV     & 1,2\\
\textbf{SDSSJ121010.13$+$334722.9} & \textbf{2.98775434(2)} & \textbf{5.0} & \textbf{6\,000$\pm$200}  & \textbf{0.415$\pm$0.010} & \textbf{RV$>$ECL} & \textbf{12}\\
SDSSJ121258.25$-$012310.2 & 8.06089(1) & 4.0 & 17\,707$\pm$35 & 0.439$\pm$0.02 & ECL    & 13,14\\
\textbf{SDSSJ122339.61$-$005631.1} & \textbf{2.1618720(3)} & \textbf{5.5} & \textbf{12\,166$\pm$114} & \textbf{0.400$\pm$0.043} & \textbf{ECL}    & \textbf{8,2}\\
SDSSJ123139.80$-$031000.3 & 5.849(9) & 4.0 & 20\,331$\pm$1173 & 0.350$\pm$0.068 & RV     & 5\\
SDSSJ124432.25$+$101710.8 & 5.468549(5) & 4.0 & 21\,535$\pm$435 & 0.400$\pm$0.026 & ECL    & 7,2\\
SDSSJ130012.49$+$190857.4 & 7.39(1) & 4.0 & 8\,657$\pm$121  & 0.960$\pm$0.103 & RV     & 5\\
SDSSJ130733.49$+$215636.7 & 5.1917311728(2) & 4.0 & --    & --    & ECL    & 8,2\\
SDSSJ134841.61$+$183410.5 & 5.962(1) & 4.0 & 15\,071$\pm$167 & 0.590$\pm$0.017 & RV     & 5\\
SDSSJ140847.14$+$295044.9 & 4.60296648(1) & 5.0 & 29\,050$\pm$484 & 0.490$\pm$0.043 & ECL    & 8,2\\
SDSSJ141536.40$+$011718.2 & 8.263939986(2) & 4.5 & 55\,995$\pm$673 & 0.564$\pm$0.014 & ECL    & 15,14\\
SDSSJ142355.06$+$240924.3 & 9.16810(4) & 5.0 & 32\,595$\pm$318 & 0.410$\pm$0.024 & ECL    & 7,2\\
SDSSJ143017.22$-$024034.1 & 4.3538160(7)& 5.0 & 10\,802$\pm$436 & 0.640$\pm$0.201 & RV     & 11,2\\
\textbf{SDSSJ143547.87$+$373338.5} & \textbf{3.015144(2)} & \textbf{5.0} & \textbf{12\,392$\pm$328} & \textbf{0.400$\pm$0.038} & \textbf{RV$>$ECL} & \textbf{3,2}\\
\textbf{SDSSJ145238.12$+$204511.9} & \textbf{2.5492327(7)} & \textbf{4.0} & \textbf{--}    & \textbf{$\geq$0.89$\pm$}    & \textbf{RV}     & \textbf{11}\\
SDSSJ145634.30$+$161137.7 & 5.498885(5) & 6.0 & 19\,416$\pm$262 & 0.370$\pm$0.016 & ECL    & 7,2\\
SDSSJ152933.25$+$002031.2 & 3.962(3) & 5.0 & 13\,986$\pm$368 & 0.385$\pm$0.032 & RV     & 1,2\\
SDSSJ154846.00$+$405728.8 & 4.4524258(4) & 6.0 & 11\,601$\pm$349 & 0.510$\pm$0.127 & RV$>$ECL & 3,2\\
SDSSJ160821.47$+$085149.9 & 9.94(3) & 6.0 & 9\,794$\pm$130  & 0.800$\pm$0.083 & RV     & 5\\
SDSSJ161113.13$+$464044.2 & 1.9768(5) & 5.0 & 10\,268$\pm$60 & 0.480$\pm$0.056 & RV$>$ELL & 5\\
SDSSJ161145.88$+$010327.8 & 7.292(6) & 6.0 & 10\,159$\pm$113 & 0.380$\pm$0.096 & RV     & 5\\
SDSSJ162552.91$+$640024.9 & 5.23771(5) & 6.0 & 8\,779$\pm$76  & 0.630$\pm$0.096 & RV     & 5\\
SDSSJ173101.49$+$623315.9 & 6.433(6) & 4.0 & 16\,159$\pm$548 & 0.410$\pm$0.054 & RV     & 5\\
SDSSJ184412.58$+$412029.4 & 5.417(1) & 6.0 & 7\,575$\pm$6  & 0.290$\pm$0.021 & RV     & 5\\
\textbf{SDSSJ211205.31$+$101427.9} & \textbf{2.2152(1)} & \textbf{6.0} & \textbf{19\,868$\pm$489} & \textbf{1.060$\pm$0.051} & \textbf{RV$>$ELL} & \textbf{5}\\
SDSSJ212320.74$+$002455.5 & 3.584(7) & 6.0 & 13\,432$\pm$928 & 0.310$\pm$0.066 & RV     & 5\\
SDSSJ213218.11$+$003158.8 & 5.334(3) & 4.0 & 16\,336$\pm$303 & 0.390$\pm$0.029 & RV     & 5\\
\textbf{SDSSJ220848.32$+$003704.6} & \textbf{2.4804(2)} & \textbf{5.0} & \textbf{--}    & \textbf{$\geq$0.33$\pm$}    & \textbf{RV}     & \textbf{11}\\
SDSSJ221616.59$+$010205.6 & 5.049(5) & 5.0 & 12\,536$\pm$1541 & 0.410$\pm$0.143 & RV     & 5\\
SDSSJ223530.61$+$142855.0 & 3.4669556448(7) & 4.0 & 21\,045$\pm$711 & 0.450$\pm$0.055 & ECL    & 8,2\\
SDSSJ224038.37$-$093541.4 & 6.254(3) & 5.0 & 13\,300$\pm$686 & 0.410$\pm$0.049 & RV     & 5\\
\textbf{SDSSJ224307.59$+$312239.1} & \textbf{2.870(6)} & \textbf{5.0} & \textbf{--} & \textbf{--} & \textbf{RV$>$ELL} & \textbf{5}\\
\hline
\end{tabular}
\end{table*}

% Don't change these lines
\bsp
\label{lastpage}
\end{document}